\documentclass[12pt]{article}
\usepackage{axodraw,bbold}

\catcode`@=12
\begin{document}
\thispagestyle{empty}
\begin{flushright} 
UCRHEP-T418\\ 
July 2006\
\end{flushright}
\vspace{0.5in}
\begin{center}
{\LARGE	\bf Suitability of $A_4$ as a Family Symmetry\\
in Grand Unification\\}
\vspace{1.5in}
{\bf Ernest Ma\\}
\vspace{0.2in}
{\sl Physics Department, University of California, Riverside, 
California 92521 \\}
\vspace{1.5in}
\end{center}

\begin{abstract}\
In the recent successful applications of the non-Abelian discrete symmetry 
$A_4$ to the tribimaximal mixing of neutrinos, lepton doublets and singlets 
do not transform in the same way.  It appears thus to be unsuitable as 
a family symmetry in grand unification.  A simple resolution of this 
dilemma is proposed.
\end{abstract}

\newpage
\baselineskip 24pt

To discover a possible underlying symmetry for the neutrino mass matrix, 
the problem of having a charged-lepton mass matrix with three very 
different eigenvalues must be faced.  A proven solution is to make use 
of a non-Abelian discrete symmetry, such as $A_4$ \cite{mr01,bmv03}. 
This is the group of even permutations of four objects, which is also 
the symmetry group of the perfect tetrahedron \cite{plato}.  It has 
12 elements and 4 irreducible representations: ${\bf 1}$, ${\bf 1'}$, 
${\bf 1''}$, and ${\bf 3}$, with the multiplication rule
\begin{equation}
{\bf 3} \times {\bf 3} = {\bf 1} + {\bf 1'} + {\bf 1''} + {\bf 3} + {\bf 3}.
\end{equation}
Specifically, for $a_{1,2,3} \sim {\bf 3}$, $b_{1,2,3} \sim {\bf 3}$, and 
$\omega = \exp (2 \pi i/3) = -1/2 + i \sqrt 3/2$,
\begin{eqnarray}
a_1 b_1 + a_2 b_2 + a_3 b_3 &\sim& {\bf 1}, \\ 
a_1 b_1 + \omega^2 a_2 b_2 + \omega a_3 b_3 &\sim& {\bf 1'}, \\ 
a_1 b_1 + \omega a_2 b_2 + \omega^2 a_3 b_3 &\sim& {\bf 1''}, \\ 
(a_2 b_3, a_3 b_1, a_1 b_2) &\sim& {\bf 3}, \\ 
(a_3 b_2, a_1 b_3, a_2 b_1) &\sim& {\bf 3}. 
\end{eqnarray}
Let
\begin{equation}
(\nu_i,l_i) \sim {\bf 3}, ~~~ l^c_i \sim {\bf 1}, {\bf 1'}, {\bf 1''}, ~~~ 
(\phi^0_i,\phi^-_i) \sim {\bf 3},
\end{equation}
then the charged-lepton mass matrix is given by
\begin{eqnarray}
{\cal M}_l &=& \pmatrix{f_1 v_1 & f_2 v_1 & f_3 v_1 \cr f_1 v_2 & f_2 \omega 
v_2 & f_3 \omega^2 v_2 \cr f_1 v_3 & f_2 \omega^2 v_3 & f_3 \omega v_3} 
\nonumber \\ 
&=& \pmatrix{v_1 & 0 & 0 \cr 0 & v_2 & 0 \cr 0 & 0 & v_3} 
\pmatrix{1 & 1 & 1 \cr 1 & \omega & \omega^2 \cr 1 & \omega^2 & \omega} 
\pmatrix{f_1 & 0 & 0 \cr 0 & f_2 & 0 \cr 0 & 0 & f_3},
\end{eqnarray}
where $\langle \phi^0_i \rangle = v_i$.  If $v_1=v_2=v_3=v$ (which breaks 
$A_4$ to the residual symmetry $Z_3$), then
\begin{equation}
{\cal M}_l = U_L \pmatrix{f_1 & 0 & 0 \cr 0 & f_2 & 0 \cr 0 & 0 & f_3} 
\sqrt 3 v,
\end{equation}
where
\begin{equation}
U_L = {1 \over \sqrt 3} \pmatrix{1 & 1 & 1 \cr 1 & \omega & \omega^2 
\cr 1 & \omega^2 & \omega}.
\end{equation}
This solves the problem of having a diagonal ${\cal M}_l$ with three 
independent eigenvalues despite the presence of an underlying symmetry.

In the neutrino sector, assuming that ${\cal M}_\nu$ is Majorana and 
allowing neutrino masses to come from Higgs triplets, it has been shown 
\cite{m04} that the form
\begin{equation}
{\cal M}_\nu = \pmatrix{a+2b & 0 & 0 \cr 0 & a-b & d \cr 0 & d & a-b}
\end{equation}
leads to tribimaximal mixing \cite{hps02} as follows.  Rotating ${\cal M}_\nu$ 
to the basis $(\nu_e,\nu_\mu,\nu_\tau)$ with $U_L$ of Eq.~(10), one obtains
\begin{eqnarray}
{\cal M}_\nu^{(e,\mu,\tau)} &=& U_L^\dagger {\cal M}_\nu U_L^* \nonumber \\ 
&=& \pmatrix{a+(2d/3) & b-(d/3) & b-(d/3) \cr b-(d/3) & b+(2d/3) & a-(d/3) 
\cr b-(d/3) & a-(d/3) & b+(2d/3)},
\end{eqnarray}
which is diagonalized exactly, resulting in
\begin{equation}
\pmatrix{\nu_e \cr \nu_\mu \cr \nu_\tau} = \pmatrix{\sqrt{2/3} & \sqrt{1/3} 
& 0 \cr -\sqrt{1/6} & \sqrt{1/3} & -\sqrt{1/2} \cr -\sqrt{1/6} & \sqrt{1/3} 
& \sqrt{1/2}} \pmatrix{\nu_1 \cr \nu_2 \cr \nu_3},
\end{equation}
where
\begin{eqnarray}
m_1 &=& a-b+d, \\ 
m_2 &=& a+2b, \\ 
m_3 &=& -a+b+d.
\end{eqnarray}
The tribimaximal pattern of Harrison, Perkins, and Scott \cite{hps02} is 
thus achieved.  However, the neutrino masses $m_{1,2,3}$ remain arbitrary. 
Nevertheless, in the special cases of $b=0$ \cite{af05} or $a=0$ 
\cite{m05-1}, predictions relating $m_{\nu_e}$ (the kinematic $\nu_e$ mass), 
$m_{ee}$ (the effective Majorana neutrino mass in neutrinoless double beta 
decay), and $\Delta m^2_{atm}$ (the mass-squared-difference measured in 
atmospheric neutrino oscillations) are obtained \cite{m05-1}.

Since $(\nu_i,l_i)$ and $l^c_i$ transform differently under $A_4$, this 
application is not compatible with grand unification in general. 
(An exception \cite{mst06} is conceivable in $SU(5)$, where $(\nu_i,l_i)$ 
belong to the ${\bf 5^*}$ representation and $l^c_i$ to the ${\bf 10}$.) 
It is thus desirable to consider instead
\begin{equation}
(\nu_i,l_i), ~l^c_i \sim {\bf 3}.
\end{equation}
In that case, the natural choice \cite{hmvv05,m05-2} of the Higgs sector is
\begin{equation}
(\phi^0_i,\phi^-_i) \sim {\bf 1}, {\bf 1'}, {\bf 1''},
\end{equation}
which exhibits an already diagonal ${\cal M}_l$.  Although the neutrino 
mixing patterm may still be consistent with data, the predictivity of 
Eq.~(11) is now lost.

The resolution of this apparent dilemma is actually very simple.  Instead 
of the Higgs assignments of Eq.~(18), let
\begin{equation}
(\phi^0_i,\phi^-_i) \sim {\bf 1}, {\bf 3},
\end{equation}
instead. Then
\begin{equation}
{\cal M}_l = \pmatrix{h_0 v_0 & h_1 v_3 & h_2 v_2 \cr h_2 v_3 & h_0 v_0 & h_1 
v_1 \cr h_1 v_2 & h_2 v_1 & h_0 v_0}.
\end{equation}
Again for $v_1=v_2=v_3=v$, it is exactly diagonalized:
\begin{equation}
{\cal M}_l = U_L \pmatrix{h_0 v_0 + (h_1+h_2)v & 0 & 0 \cr 0 & h_0 v_0 + 
(h_1 \omega + h_2 \omega^2)v & 0 \cr 0 & 0 & h_0 v_0 + (h_1 \omega^2 + 
h_2 \omega)v} U_R^\dagger,
\end{equation}
where $U_L = U_R$ is given again by Eq.~(10).  This means that Eq.~(11) 
for ${\cal M}_\nu$ again predicts tribimaximal mixing as desired. 
The obstacle to having a grand-unified model within this context is 
removed.

In the quark sector, ${\cal M}_u$ and ${\cal M}_d$ are also of the form of 
Eq.~(20), thus predicting $V_{CKM} = 1$ in the limit of $v_1=v_2=v_3=v$ 
as noted before \cite{bmv03}.  This is not such a bad first approximation 
and it may be possible to recover a realistic $V_{CKM}$ from explicit soft 
$A_4$ breaking terms.

In the context of the discrete symmetry $\Delta(27)$ which has recently been 
proposed \cite{mvkr06,m06}, the analogous ${\bf 3} \times {\bf 3} \times 
{\bf 3}$ decomposition has 3 invariants, resulting in the form
\begin{equation}
{\cal M}_l = \pmatrix{h_0 v_1 & h_1 v_3 & h_2 v_2 \cr h_2 v_3 & h_0 v_2 & h_1 
v_1 \cr h_1 v_2 & h_2 v_1 & h_0 v_3},
\end{equation}
which has the same limit as Eq.~(20) for $v_1=v_2=v_3=v$.  However, Eq.~(11) 
for ${\cal M}_\nu$ is not as easy to obtain as in $A_4$.

In conclusion, thanks to Eq.~(21), it has been shown that $A_4$ is a 
suitable candidate family symmetry in grand unfication, allowing all quarks 
and leptons to transform according to its ${\bf 3}$ representation, and 
predicting the tribimaximal mixing of neutrinos.

This work was supported in part by the U.~S.~Department of Energy under 
Grant No.~DE-FG03-94ER40837.

\newpage
\bibliographystyle{unsrt}

\end{document}